\documentstyle[12pt]{article}

\def\labelmark{}
\def\void{}

{\ifx\void\labelname\def\junk{\end{displaymath}}
\else\def\junk{\end{eqnarray}}\fi\junk\labelmark\def\labelname{}}

\newcommand{\bra}{\begin{array}}
\newcommand{\era}{\end{array}}
\newcommand{\beq}{\begin{equation}}
\newcommand{\eeq}{\end{equation}}
\newcommand{\bqn}{\begin{eqnarray}}
\newcommand{\eqn}{\end{eqnarray}}

\def\BC{\bb C}
\def\_\BC{\bbi C}

\newcommand{\om}{\omega}

\newcommand{\la}{\lambda}

\newcommand{\be}{\beta}

\newcommand{\te}{\theta}

\newcommand{\pa}{\partial}
\newcommand{\al}{\alpha}

\newcommand{\na}{\nabla}

\newcommand{\st}{\star}
\newcommand{\ti}{\tilde}
\newcommand{\da}{\dagger}

\newcommand{\PR}[1]{ { Phys.~Rev.} { #1}}

\newcommand{\EPJ}[1]{ { Eur.~Phys.~J.} { #1}}

\newcommand{\JSP}[1]{ { J.~Stat.~Phys.} { #1}}

\begin{document}
\begin{titlepage}
\setcounter{page}{1}
\renewcommand{\thefootnote}{\fnsymbol{footnote}}


\begin{center}
{\Large\bf Landau Diamagnetism in  Noncommutative Space
and the Nonextensive Thermodynamics of Tsallis}

\vspace{17mm}

{\bf{\"{O}mer F. Dayi}$^{a,b}$ 
\footnote{E-mail: dayi@itu.edu.tr and dayi@gursey.gov.tr.}}
\,{and}\,
{ \bf{Ahmed Jellal}$^{a}$ 
\footnote{E-mail: jellal@gursey.gov.tr.}}\\
\vspace{5mm}

\end{center}

\noindent
{\em $^{a}$ {\it Feza G\"{u}rsey Institute, P.O. Box 6, 81220,
\c{C}engelk\"{o}y, Istanbul, Turkey. } }

\vspace{3mm}

\noindent
{\em $^{b}${\it Physics Department, Faculty of Science and
Letters, Istanbul Technical University,\\
80626 Maslak--Istanbul, Turkey.} } 

\vspace{5mm}

\begin{abstract}
We consider the behavior of
electrons in an external uniform magnetic field $\vec{B}$
where the space coordinates 
perpendicular to $\vec{B}$ are taken as noncommuting.
This results in a generalization of  standard thermodynamics.
Calculating the susceptibility, we find that  the usual Landau diamagnetism 
is modified. We also compute
the susceptibility 
according to the nonextensive statistics
of Tsallis for $(1-q) \ll 1$, in terms of the factorization approach. 
Two methods agree under certain conditions.

\end{abstract}

\end{titlepage}

\newpage
\section{Introduction}

The natural appearance of noncommutativity in string theories~\cite{cds} 
and
the observation that ordinary and noncommutative gauge theories
are equivalent~\cite{sw} have increasingly led to  attempts  to  study
physical problems in noncommutative spaces. 
Apparently, formulation of quantum mechanics in noncommutative 
spaces~\cite{mez},~\cite{oth}
may provide some hints for
understanding the physical consequences of  noncommuting
space coordinates. 

Although noncommuting coordinates are operators 
even at the classical level, one can treat them as commuting by
replacing operator products with $*$--products. This approach was developed
in order 
to formulate quantum mechanics without appealing to operators~(\cite{bay}
and the references therein). However, it is 
more appropriate to perform calculations
with noncommutative coordinates without introducing new machinery. 
This approach allows 
us to generalize classical as well as 
quantum mechanics without altering their main physical interpretations and to 
recover the usual results when noncommutativity is switched off.

There exist some physical systems which do not obey the rules of 
the standard thermodynamics (some of them are mentioned in \cite{trev}). 
One may hope to find a remedy for these anomalies by generalizing
Boltzmann--Gibbs statistics
by adopting noncommutative coordinates.
At this point, we cannot offer a general answer
as to whether this is always possible.
Nevertheless, we can obtain  insights by studying a specific example.

Calculation of susceptibility for  electrons moving in an external 
uniform  magnetic field leads to Landau diamagnetism~\cite{lan},
which is a consequence  of quantization. 
We wish to study this system,
which   may provide some evidence on whether   
the generalization of the usual thermodynamics due to
noncommutativity of space coordinates 
is useful 
when Boltzmann--Gibbs statistics is not applicable.
One can choose a gauge such that the 
motion in the direction of the magnetic field remains intact. 
It is then convenient to
calculate  the susceptibility in a setting where the coordinates which are 
perpendicular to the external magnetic field are taken as noncommuting,
but the one in the direction of the external magnetic field is kept
commuting. 

On the other hand  Tsallis~\cite{tsa},\cite{trev} proposed
a nonextensive statistics,
which  is a candidate 
for curing some of 
the problems that appear in  standard thermodynamics.
It is therefore interesting to investigate whether
the statistics resulting from  noncommutativity of the coordinates
is somehow  connected to the 
nonextensive statistics of Tsallis. 

After a brief review of the quantization  of an electron moving 
in an external uniform magnetic field and the related coherent states,
we calculate the partition function, the magnetization  and 
the susceptibility, taking
the coordinates  perpendicular to the magnetic field  as
noncommuting.

Although
Landau diamagnetism according to Tsallis
nonextensive statistics was studied
in~\cite{otbd}, the method adopted therein
is not suitable for our purposes.
We therefore,
present  another method to calculate the susceptibility 
according to  the 
nonextensive statistics of Tsallis 
by retaining the leading terms
in  $(q- 1)\ll 1.$  
These two
methods
yield different results. 

We then show that
in the high temperature limit 
the susceptibility obtained in the
noncommuting space and the one based on Tsallis 
nonextensive statistics
agree if their parameters satisfy a certain relation.

\section{Preliminaries}

An electron moving in the external uniform magnetic field ${\vec 
B}=\vec\na\times\vec A,$ neglecting spin, is 
described by the Hamiltonian
\beq
\label{fh}
H = {1\over 2m}({\vec p}+{e\over c}{\vec A})^2.
\eeq
We let the magnetic field ${\vec B}$ be in the $z$-direction by
choosing the symmetric gauge:
\beq
\label{gco}
{\vec A}=(-{B\over 2}y,{B\over 2}x,0).
\eeq
After imposing the usual canonical commutation relations between the 
coordinates $\vec{r}$ and the momenta $\vec{p},$ the operators
\beq
\label{oa}
a= \frac{1}{\sqrt{2m \hbar \omega}}\left[
\left( p_x-\frac{m\omega}{2} y\right) 
-i\left( p_y+\frac{m\omega}{2} x\right) 
\right] ,
\eeq
and its hermitian conjugate $a^\dagger$ 
can be shown to satisfy
\beq
[a,a^{\da}]=1,
\eeq
where $\om={eB\over mc}$ is the cyclotron frequency.

In the $z$--direction
the motion is free, however,
the appropriate  Hamiltonian for the motion in the $(x,y)$--plane
becomes
\beq
\label{ham}
H_\perp = \hbar\om(a^{\da}a+{1\over 2}).
\eeq
Eigenstates of (\ref{ham}) can  simultaneously
be taken to be 
eigenstates of angular momentum in the $z$--direction.
Although, one can construct coherent states reflecting this fact~\cite{fk}, 
for our purposes it is sufficient to consider the
normalized coherent states $|\alpha >$ defined as
\beq
\label{oc}
|\al>\equiv \exp\Big(-{|\al|^2\over 4l^2}-i\sqrt{m\om\over 
2\hbar}a^{\da}\Big)
|0> ,
\eeq
where $l=({\hbar\over m\om})^{1\over 2}$ is the magnetic length and $\al$
is a complex parameter. As usual the ground state $|0>$ is defined to
satisfy
$a|0> = 0 ,$
and $|\al >$ are eigenstates of the annihilation operator:
\beq
\label{orc}
a|\al>=\Big({\hbar\over 2m\om}\Big)^{1\over 2}{\al\over il^2}|\al>,
\eeq
These states were used in \cite{fk}  to calculate the
susceptibility for a cylindrical body  which results in the usual Landau 
diamagnetism.

\section{Landau diamagnetism in noncommutative space}

Because of choosing the symmetric gauge (\ref{gco}),
the motion  along the $z$--direction is free and
the motion in the plane perpendicular to the magnetic field
$\vec{B}$ is described by (\ref{ham}).
Let us keep the $z$ coordinate commuting, but take the 
coordinates perpendicular to the magnetic field $\vec{B}$ 
as  noncommuting 
\beq
\label{com}
[\hat{x},\hat{y}]=i\te,
\eeq 
where the constant $\te$ is  real.
Noncommutativity can be imposed 
by treating the coordinates as
commuting, but
introducing the star product 
\beq
\label{star}
\st\equiv\exp \frac{i\te}{2} \Big(
{\stackrel\leftarrow\pa}_{x} {\stackrel\rightarrow\pa}_{y}
-{\stackrel\leftarrow\pa}_{y}{\stackrel\rightarrow\pa}_{x}
\Big) .
\eeq 
Now, we deal with  the commutative coordinates $x$ and $y$ but 
 replace the products  with the star product 
(\ref{star}). For example, 
instead of the commutator  (\ref{com}) one introduces the Moyal bracket
\beq
x\st y-y\st x=i\te .
\eeq 
We quantize  this system in terms of the standard canonical quantization
by establishing the usual canonical commutation relations
${[r_i,p_j]}=i\hbar \delta_{ij} .$
However, eigenvalue equations should be modified.
According to this receipt,  in the symmetric gauge (\ref{gco}), the
transverse part of the Hamiltonian (\ref{fh}) acting 
on an arbitrary function  $\Psi (\vec{r},t)$ yields~\cite{mez}
\beq
H_\perp \st \Psi (\vec{r},t) =
\frac{1}{2m}\left[
\left( p_x -\frac{eB}{2c} y \right)^2 +
\left( p_y +\frac{eB}{2c} x \right)^2 \right]\st\Psi (\vec{r},t)
\equiv H_{nc} \Psi (\vec{r},t),
\eeq
where, in terms of $\kappa =\frac{eB\theta}{4c},$
we  defined
\beq
H_{nc} =
\frac{1}{2m}\left[
\left( (1+\kappa )p_x -\frac{eB}{2c} y \right)^2 +
\left( (1+\kappa )p_y +\frac{eB}{2c} x \right)^2 \right].
\eeq
We obtain the appropriate creation and annihilation operators for this
system, ${\ti a}^{\da}$ and ${\ti a},$  from (\ref{oa}) 
by replacing
$\omega $ with the modified  cyclotron frequency $\ti{\omega}:$ 
\beq
\ti{\om}\equiv  (1+\kappa ) \omega .
\eeq
They satisfy 
the usual commutation relation 
$$
[{\ti a},{\ti a}^{\da}]=1.
$$ 
The modified transverse Hamiltonian $H_{nc}$
can be written as
\beq
H_{nc}=\hbar\ti{\om}({\ti a}^{\da}{\ti a}+{1\over 2}).
\eeq
Reminiscent of noncommutativity is only in the frequency 
$\tilde{\omega} .$
In fact we construct the
coherent states $|\ti{\al}>$ of the noncommutative case 
from  $|{\al}>$  
by the replacement
$a\rightarrow \ti{a},$ $a^\dagger \rightarrow {\ti{a}}^\dagger ,$ 
$\om \rightarrow \ti{\om}$
in (\ref{oc}). Obviously, they  satisfy
\beq
{\ti a}|\ti{\al}>=\Big({\hbar\over 2m{\ti\om}}\Big)^{1\over 2}
{\al\over i\ti{l}^2}|\ti{\al}> ,
\eeq
where $\ti{l}=({\hbar\over m\ti{\om}})^{1\over 2}.$

The creation and annihilation operators ${\tilde{a}}^\dagger$ and
$\tilde{a}$
are the usual ones appearing in
quantum mechanics.
Thus, the total Hamiltonian
\[
H_t=\frac{p_z^2}{2m} +H_{nc},
\]
can be used to define
the partition function in noncommutative 
coordinates in the standard way as
\beq
Z_{nc}= {\textbf{Tr}}\ e^{-\be H_{t}} ,
\eeq
where $\beta =1/kT. $ 

Now, by following the arguments of~\cite{fk}, the partition 
function $Z_{nc}$ 
for a cylindrical body of volume $V$
can be written as
\beq
Z_{nc}= {V\over \la^3}\;{\be\hbar \ti{\om} \over 2}\;
{e^{-{{\be\hbar\ti{\om}\over 2}}}\over 2{\ti{l}}^2}\int_0^{\infty}
d^2{\al}<{\ti\al}|e^{-\hbar\ti{\om}\be{\ti a}\da{\ti a}}|{\ti\al}>,
\eeq
where $\la=\Big({2\pi\hbar^2\be
\over m}\Big)^{1\over 2}$ is the thermal wavelength. 
By performing the calculation
it leads to
\beq
Z_{nc}= {V\over \la^3}\;{\be\hbar\om\over 2}\;
{(1+\kappa )\over \sinh\Big({\be\hbar{\om}\over 2}(1+\kappa )\Big)},
\eeq
where we emphasized the $\kappa$ dependence.
We should adopt the standard definitions for the free energy and the
magnetization:
\beq
F_{nc}= -{n\over\be}\ln Z_{nc},\  
M_{nc}= -{\pa F_{nc}\over\pa B},
\eeq
where $n$ is the total number of particles.
Thus, we compute the magnetization as
\beq
M_{nc}={n\hbar e\over mc}\;(1+2\kappa )
\Bigg[{1\over \be\hbar\om (1+\kappa )}-{1\over 2}
\coth\Big({\be\hbar \om \over 2}(1+\kappa )\Big)\Bigg].
\eeq
The susceptibility   defined by
\beq
\chi_{nc}= {1\over n}{\pa M_{nc}\over\pa B},
\eeq
can be calculated to yield
\beq
\bra{l}
\chi_{nc}={\hbar e\over mc}\;2\kappa 
\Bigg[{1\over \be\hbar\om (1+\kappa )}-{1\over 2}
\coth\Big({\be\hbar \om \over 2}(1+\kappa )\Big)\Bigg]\\
\qquad\;\;-\Big({\hbar e\over mc}\Big)^2\be\;(1+2\kappa )^2\;
\Bigg[{1\over \Big(\be\hbar\om(1+\kappa )\Big)^2}+{1\over 4}
\Big[1-\coth^2\Big({\be\hbar{\om}\over 2}(1+\kappa )\Big)\Big]\Bigg] .
\era
\eeq
In the high temperature limit, $\be\ll1,$ we get
\beq
\label{htt}
\chi_{nc}=\chi_{L}\Big[1+
6\kappa + 6\kappa^2
\Bigg],
\eeq
where the usual Landau diamagnetism is
$$\chi_{L}=-{1\over 3}\;\Big({\hbar e\over 2mc}\Big)^2\be .$$

Noncommutativity arising in string theories is due to a background field
strength~\cite{cds}. This leads to the fact that the related  noncommutativity
parameter is positive. However, when we introduce the noncommutativity in
an ad hoc manner in terms of the commutator (\ref{com}), there is no
restriction on
the sign of  $\te .$ In spite of that, for the consistency of our
formalism 
$\tilde{\omega} $
should be positive, which yields the condition
$\kappa =\frac{eB\theta}{4c} > -1.$ For these values of $\kappa$ 
(or $\te $) the system is diamagnetic, except for the values 
$-0.8< \kappa < -0.2,$ where 
$\chi_{nc}$ becomes positive.
This is an interesting fact
which deserves a more detailed study.

\section{Landau diamagnetism according to the nonextensive
         statistics of Tsallis}

To deal with the systems which do not obey the rules of
the usual Boltzmann-Gibbs statistics 
Tsallis  proposed a nonextensive generalization of
thermodynamics~\cite{tsa},\cite{trev}. In this formalism
internal energy can be constrained in three different ways
which result in different definitions of partition
function~\cite{ct},\cite{tmp}. We take the definition
of the partition function to be
\beq
Z_q=  \Big [1-(1-q)\be \sum_nE_n \Big ]^{1\over 1-q} ,
\eeq
where the parameter $q$ is a real number and $E_n$ are energy eigenvalues.
This definition possesses some undesirable properties and the most
satisfactory definition which we denote as $\ti{Z}_q(\beta )$ is given in
terms of normalized $q$--expectation values~\cite{tmp}. Nevertheless,
there exists
a map~\cite{tmp} $\beta \rightarrow \beta^\prime$ such that 
$\ti{Z}_q(\beta ) =Z_q(\beta^\prime ).$ Moreover, $\beta^\prime$ is an 
increasing function of $\beta $~\cite{lp}, 
so that the high temperature
limit can equivalently be taken as $\beta^\prime \ll 1.$

The free energy $F_q$ is also modified:
\beq
\label{f-q}
F_q=\;-{1\over\be}\;{Z_q^{1-q}-1\over 1-q},
\eeq
but the magnetization $M_q$ and the susceptibility $\chi_q$
are still given by
\beq
\label{mc}
M_q=\;-{\pa F_q\over\pa B},\qquad \chi_q=\;{1\over n}\;{\pa M_q\over\pa B}.
\eeq

We wish to investigate whether
 the generalized statistics obtained due to
noncommutativity of space and the nonextensive statistics of Tsallis
possess some common features. Obviously, 
we can have insights by
studying the specific example of the previous section, namely 
electrons in the
uniform magnetic field $\vec{B}$
according to the nonextensive statistics of Tsallis
and then comparing their results.

Calculation of generalized partition functions exactly is a hard 
task~\cite{cu} and it is not always possible to achieve. We use the
approximation which is known as the factorization approach~\cite{bdg}. 
It is valid for
high and low temperatures without any restriction on $q$
and for small number of states for $(1-q)\ll 1$~\cite{ppp}. However, its
validity can be enlarged to other values of $q$ as far as some other
conditions are satisfied~\cite{wm},\cite{baa}.

Our aim is to calculate the generalized partition function $Z_q$
for a cylindrical body of length 
$L$ and radius $R,$ which is  oriented along the $z$-direction,
using the Hamiltonian~(\ref{fh}).  The transverse part of the
Hamiltonian, by choosing the symmetric
gauge~(\ref{gco}), is given as in (\ref{ham}).
By using the factorization approach
the generalized partition function $Z_q$ can be 
written as
\beq
\label{zq}
Z_q\simeq \;{L\over \hbar}\;\Big({ m\over 2\pi\be}\Big)^{1\over 2}\;
\Big [1-{\be\hbar\om\over 2}(1-q)\Big ]^{1\over 1-q}\;Z_{\perp},
\eeq
where $Z_\perp $ is taken as
\beq
\label{zper}
Z_{\perp}={R^2\over 4\pi l^4}\int d^2\al<\al|
\Big [1-(1-q)\be\hbar\om a^{\da}a\Big ]^{1\over 1-q}|\al>.
\eeq
The coherent states $|\alpha > $ are the eigenstates of $a$
satisfying  (\ref{orc}). Thus,
if we  write the operators appearing
in (\ref{zper}) in normal ordered form
the computation will become manageable.
Fortunately, it is known that any function $f$ of
 $a^{\da}a$  can be written in the normal ordered form as~\cite{lou}
\beq
\label{fun}
f(a^{\da}a)=\sum_{r=0}^{\infty}\sum_{s=0}^{r}{(-1)^sf(r-s)\over (r-s)!s!} 
a^{{\da}r}a^r.
\eeq
Actually, 
for an exponential function it yields
\beq
\label{vnf}
e^{\xi a^{\da}a}=\sum_{r=0}^{\infty}{(e^{\xi}-1)^r\over r!} a^{{\da}r}a^r,
\eeq
where $\xi$ is a constant. By making use of
(\ref{fun}) and (\ref{orc})  we 
can write (\ref{zper}) as 
\beq
Z_{\perp}={R^2\over 4\pi l^4}\int d^2\al\sum_{r=0}^{\infty}
C_r\Bigg({|\al|^2\over 2l^2}\Bigg)^r ,
\eeq
where  the coefficients $C_r$ are given by
\beq
C_r=\sum_{s=0}^{r}{(-1)^s\over (r-s)!s!}
\Big [1-(1-q)\be\hbar\om (r-s)\Big ]^{1\over 1-q}.
\eeq
These coefficients can be expanded 
in  $(1-q)$ as
\beq
C_r=\sum_{s=0}^{r}{(-1)^s\over (r-s)!s!}\sum_{n=0}^{\infty}
{(-1)^n\over (1-q)^n n!}\Bigg(\sum_{k=1}^{\infty}{(1-q)^k\over k}
(\be\hbar\om)^k (r-s)^k\Bigg).
\eeq
The terms to  second order in   $(1-q)$ can be
evaluated explicitly 
\begin{eqnarray*}
C_r & = & \sum_{s=0}^{r}{(-1)^s\over (r-s)!s!}\sum_{n=0}^{\infty}
{(-1)^n\over n!}\Bigg\{(\be\hbar\om)^n+(1-q){n\over 2}(r-s)^{n+1}
(\be\hbar\om)^{n+1}\\
 & &  \qquad\;+(1-q)^2\Big [{n(n-1)\over 8}+{n\over 3}\Big ]
(r-s)^{n+2}(\be\hbar\om)^{n+2}+...\Bigg\}.
\end{eqnarray*}  
We retain  the terms  to first order in  $(1-q)$ and
utilize  (\ref{vnf}) to obtain  the compact form
\beq
C_r\simeq {(e^{-\be\hbar\om}-1)^r\over r!}-{(1-q)\over 2}\be^2
{\pa^2\over\pa\be^2}\Bigg[ {(e^{-\be\hbar\om}-1)^r\over r!}\Bigg ].
\eeq
Now, we   write $Z_\perp$ as
\beq
Z_{\perp}={R^2  4\pi l^4}\Bigg(1-{(1-q)\over 2}\be^2
{\pa^2\over\pa\be^2}\Bigg)\int d^2\al
\exp\Bigg[-{|\al|^2\over 2l^2}\Big(1-e^{-\be\hbar\om}\Big)\Bigg],
\eeq
which can be evaluated to yield
\beq
\label{lzp}
Z_{\perp}={R^2\over 2l^2}\Bigg(1-{(1-q)\over 2}\be^2
{\pa^2\over\pa\be^2}\Bigg)\Bigg({1\over 1-e^{-\be\hbar\om}}\Bigg).
\eeq
By keeping the dominant terms in 
$(1-q),$ 
the first factor appearing in (\ref{zq})
can be written as
\beq
\label{of}
\Big [1-{\be\hbar\om\over 2}(1-q)\Big ]^{1\over 1-q}\;\;\simeq\;\;
e^{-{\be\hbar\om\over 2}}-{(1-q)\over 2}\Big({\be\hbar\om\over 2}\Big)^2
e^{-{\be\hbar\om\over 2}}.
\eeq
Substitution of the
approximated values 
(\ref{lzp}) and (\ref{of}) into~(\ref{zq}) yields
\beq
Z_q\simeq \;{V\over \la^3}\;
{\be\hbar\om\over 2}\;\Bigg\{{1\over\sinh(\be\hbar\om/2)}
-{(1-q)\over 2}\;\Big[{(\be\hbar\om/2)^2\over\sinh(\be\hbar\om/2)}
+e^{-\be\hbar\om\over 2}\be^2{\pa^2\over\pa\be^2}
(1-e^{-\be\hbar\om})^{-1}\Big]\Bigg\}.
\eeq
We can write it in the compact form
\beq
\label{z-q}
Z_q\simeq \;Z_{L}\;\Big[1-\frac{(1-q)}{2}\;\Sigma(\be,\om)\Big],
\eeq
where $Z_L$ is the partition function of 
free electrons in the magnetic field $\vec{B}$
according to
the standard thermodynamics
\beq
Z_L=\;{V\over \la^3}\;
{\be\hbar\om\over 2}\;{1\over\sinh(\be\hbar\om/2)},
\eeq
and the first order 
modification is
\beq
\Sigma(\be,\om)=(\be\hbar\om)^2\;\Bigg[ {1\over 4}+{1\over 2}
e^{-\be\hbar\om\over 2}{\cosh(\be\hbar\om/2)\over
\sinh^2(\be\hbar\om/2)}\Bigg].
\eeq

Because of dealing with the values of $q$ such that $(1-q)\ll 1,$ 
we retain the terms to  first order  in $(1-q)$ for the free
energy $F_q$ 
\beq
F_q  \simeq\; -{1\over \be}
\Bigg[\ln Z_L-{(1-q)\over 2}\;\Sigma(\be,\om)-
{(1-q)\over 2}(\ln Z_L)^2\Bigg],
\eeq
after inserting (\ref{z-q})  into (\ref{f-q}).
It is now straightforward to calculate the magnetization $M_q$
and the susceptibility $\chi_q$ of the nonextensive case by making use of
the
definitions (\ref{mc}), which lead to 
\begin{eqnarray}
M_q & \simeq\; & M_L-{(1-q)\over 2}\;{1\over\be}\;{\pa \Sigma\over\pa B}+
(1-q)M_L\ln Z_L,\\
\chi_q & \simeq\; & \chi_L-{(1-q)\over 2}\Bigg[{1\over\be}\;{\pa^2 \Sigma\over\pa 
B^2}
+{2\be\over n^2}M_L^2+2\chi_L\ln Z_L\Bigg], \label{chiq}
\end{eqnarray}
where $M_L=\;{\pa\ln Z_L\over\pa B}$ and 
$\chi_L=\;{1\over n}\;{\pa M_L\over\pa B}$ are the
magnetization and the susceptibility according to  the standard extensive
formalism
which can be 
approximated in the high temperature limit, $\beta \ll 1,$ as 
\begin{eqnarray}
M_L & \simeq\; & -{1\over 6}\;{ne\hbar^2\om\over mc} \be,\\
\chi_L & \simeq\; & -{1\over 3}\;\Big({\hbar e\over 2mc}\Big)^2\;\be .
\end{eqnarray}
Moreover, the leading terms of $\Sigma(\be ,\omega )$
for high temperatures are
\beq
\Sigma(\be,\om)\simeq\;2-\be\hbar\om+{7\over 3}\Big(
{\be\hbar\om\over 2}\Big)^2.
\eeq
For  $\beta \ll 1,$ 
the third term of (\ref{chiq})
behaves like $\be^3$ so that it should be ignored with respect to
the second term behaving like $\be$ . 
We then conclude that
according to the nonextensive formalism of Tsallis
the standard Landau diamagnetism 
is modified as
\beq
\label{lap}
\chi_q\simeq\;\chi_L\;\Big[1+(1-q) (7+ \ln Z_L )\Big],
\eeq
for  $\beta\ll 1$ as well as retaining only 
the terms to first order  in 
$(1-q).$

\section{Conclusions}

The modified susceptibility in noncommuting coordinates (\ref{htt}) for 
$\beta\ll 1$ and the one obtained according to the 
nonextensive formalism of Tsallis (\ref{lap})  for $\beta\ll 1$
and $(1-q)\ll 1,$  in terms of the factorization approach,
coincide  if their parameters are related as
\beq
\kappa +\kappa^2
=(1-q)\left(\frac {7+ \ln Z_L}{6} \right).
\eeq
We recall that $\kappa$ is given in terms of the noncommutativity
parameter $\theta$ as  $\kappa =\frac{eB\theta}{4c}.$
Thus, the generalization of the standard thermodynamics obtained
due to noncommutativity of coordinates covers,  in certain limits,
the nonextensive
formalism of Tsallis for Landau diamagnetism. We would like to emphasize
that 
this result is obtained in terms of some approximations and it is far from
being an exact result. 

Obviously, it may happen that some deformations of quantum
mechanical systems other than letting the coordinates be noncommuting 
may lead to similar relations. Nevertheless, this does not alter the fact
that our results indicate that the generalized statistics due to
noncommutative
coordinates is a candidate to deal with the systems 
which do not obey the rules of the standard thermodynamics.
However, it is not guaranteed that the results obtained in this specific
example can be extrapolated to  other physical systems.
To have a better understanding of the roles of noncommuting coordinates in
generalization of the standard thermodynamics one should study  more
physical systems in noncommuting coordinates.

\vspace{5mm}
\begin{center}
{\bf Acknowledgements}
\end{center}

\noindent
For very fruitful discussions 
we are indebted to A.~Erzan and U.~T{\i}rnakl{\i}.
We would like to thank E.~\.{I}n\"on\"u 
and C. Sa\c{c}l{\i}o\u{g}lu for their interest. A.J. also 
thanks  J.~Brodie and H.~Saidi  for their encouragement.

\end{document}